\documentclass[conference]{IEEEtran}
\IEEEoverridecommandlockouts
\usepackage{cite}
\usepackage{amsmath,amssymb,amsfonts}
\usepackage{algorithmic}
\usepackage{graphicx}
\usepackage{textcomp}
\usepackage{xcolor}
\usepackage{float}
\usepackage{caption, subcaption}
\def\BibTeX{{\rm B\kern-.05em{\sc i\kern-.025em b}\kern-.08em
    T\kern-.1667em\lower.7ex\hbox{E}\kern-.125emX}}

\usepackage[english]{babel}
\emergencystretch=\maxdimen
\begin{document}

\title{A novel approach towards the classification of Bone Fracture from Musculoskeletal Radiography images using Attention Based Transfer Learning\\
}

\author{\IEEEauthorblockN {Sayeda Sanzida Ferdous Ruhi}
\IEEEauthorblockA{\textit{Dept. of CSE} \\
\textit{Stamford University Bangladesh}\\
Dhaka, Bangladesh \\
ssfruhi@gmail.com}
\and
\IEEEauthorblockN{Fokrun Nahar}
\IEEEauthorblockA{\textit{Dept. of CSE} \\
\textit{Stamford University Bangladesh}\\
Dhaka, Bangladesh \\
fokhrunnaharemu@gmail.com}
\and
\IEEEauthorblockN{Adnan Ferdous Ashrafi}
\IEEEauthorblockA{\textit{Dept. of CSE} \\
\textit{Stamford University Bangladesh}\\
Dhaka, Bangladesh \\
adnan@stamforduniversity.edu.bd}}

\maketitle

\begin{abstract}

Computer-aided diagnosis (CAD) is today considered a vital tool in the field of biological image categorization, segmentation, and other related tasks. The current breakthrough in computer vision algorithms and deep learning approaches has substantially enhanced the effectiveness and precision of apps built to recognize and locate regions of interest inside medical photographs. Among the different disciplines of medical image analysis, bone fracture detection, and classification have exhibited exceptional potential. Although numerous imaging modalities are applied in medical diagnostics, X-rays are particularly significant in this sector due to their broad availability, ease of use, and extensive information extraction capabilities. This research studies bone fracture categorization using the FracAtlas dataset, which comprises 4,083 musculoskeletal radiography pictures. Given the transformational development in transfer learning, particularly its efficacy in medical image processing, we deploy an attention-based transfer learning model to detect bone fractures in X-ray scans. Though the popular InceptionV3 and DenseNet121 deep learning models have been widely used, they still have the potential to be employed in crucial jobs. In this research, alongside transfer learning, a separate attention mechanism is also applied to boost the capabilities of transfer learning techniques. Through rigorous optimization, our model achieves a state-of-the-art accuracy of more than 90\% in fracture classification. This work contributes to the expanding corpus of research focused on the application of transfer learning to medical imaging, notably in the context of X-ray processing, and emphasizes the promise for additional exploration in this domain.
\end{abstract}


\section{Introduction}
Bone fractures are one of the most common injuries nowadays. Every year, 2.7 million fractures occur across the EU6 nations, France, Germany, Italy, Spain, Sweden, and the UK \cite{data}. A bone fracture occurs when there is a break or cracks in a bone, resulting in a disruption of its normal structure and function. Fracture creates hazards to one's daily life. and sometimes, it is proven life-threatening if the fracture occurs in the skull or in any other parts of the body that are directly connected to the cause of death. That's the reason why detecting fractures is a vital job. It is one of the most challenging works in bio-medical subjects. Detecting fractures means detecting each single abnormality in bones and it requires more focus. By identifying fractures, we can find out how damaged the bones are and what measures need to be taken to overcome the damage.


\begin{figure*}[ht]
\begin{center}
\begin{subfigure}[b]{0.49\textwidth}
\includegraphics[width=\textwidth]{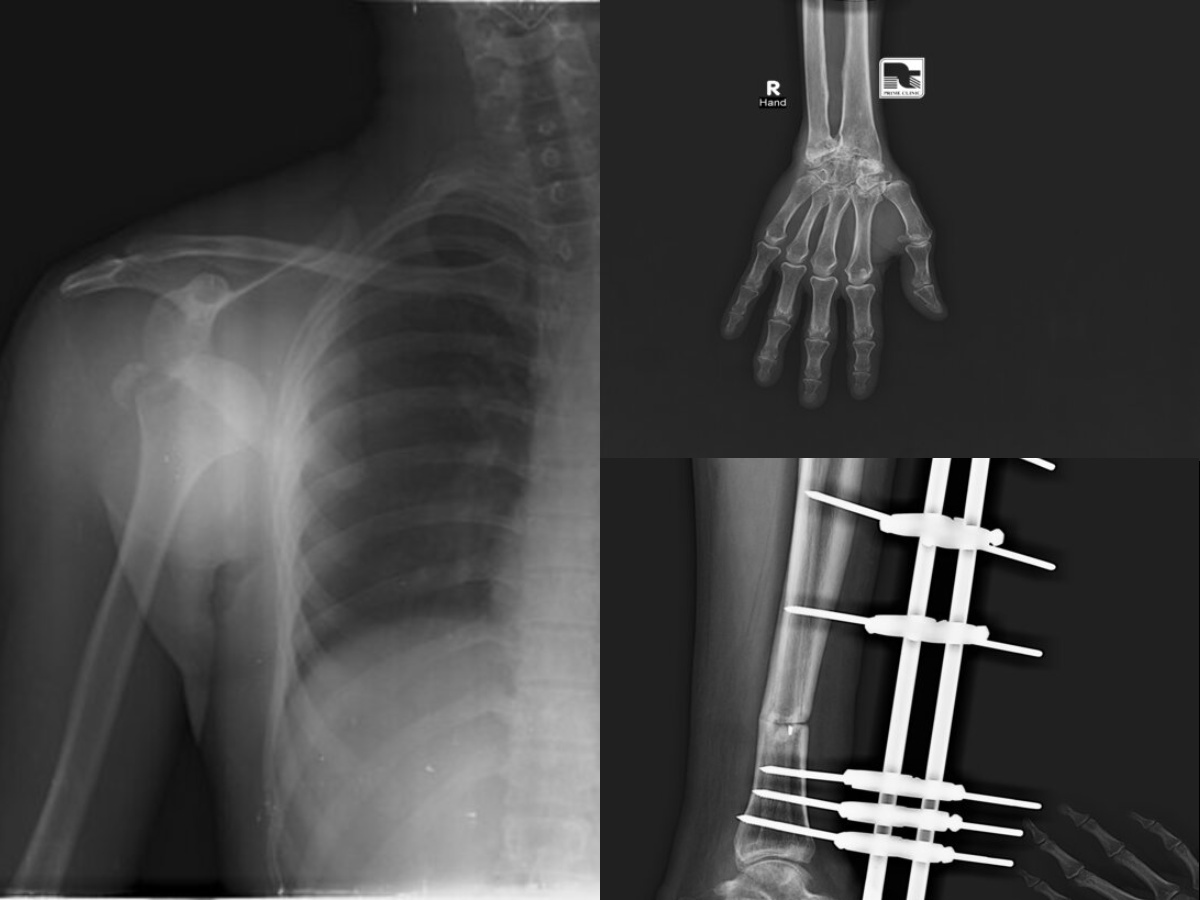}

\caption{Image showing fracture in different parts of body}
\end{subfigure}
\hfill
\begin{subfigure}[b]{0.49\textwidth}
    \begin{center}
\includegraphics[width=\textwidth]{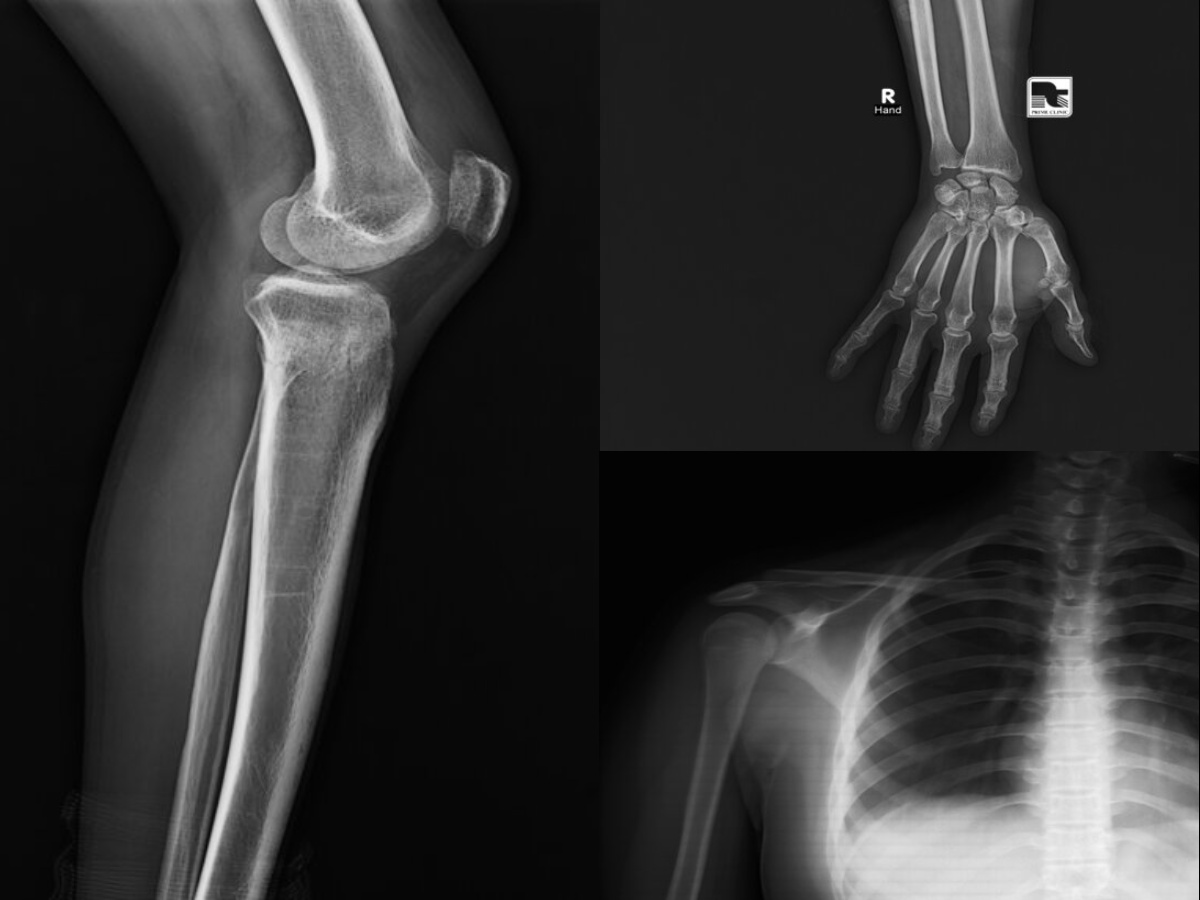}
\end{center}
\caption{Image showing no fracture in different parts of body}
\end{subfigure}
\end{center}
\caption{Sample images from the FracAtlas Dataset}
\label{fig:dataset}
\end{figure*}

The employment of AI models in the diagnosis and categorization of fractures constitutes a radical change in the field of orthopedic treatment. These models introduce an unparalleled degree of efficiency, utilizing their powerful computing capabilities to evaluate huge amounts of medical imaging data with amazing speed and precision. This capacity not only expedites the diagnostic procedure but also increases the accuracy of fracture classification, ultimately improving patient outcomes. The precision supplied by AI models is particularly remarkable since it considerably minimizes the chance of human mistakes, which can be frequent in traditional diagnostic approaches.

Moreover, the incorporation of AI with medical imaging aids in the treatment of complex situations where minor fracture patterns may be tough to recognize. By combining advanced algorithms, AI models may discover these details that could otherwise be unnoticed by human sight. This degree of detail emphasizes the potential of AI to transform diagnostic processes in orthopedics, delivering an accurate and consistent technique for fracture categorization. Various types of AI tools are available for this purpose, including computer vision libraries that specialize in image processing and feature extraction, machine learning frameworks that enable the development of predictive models, and pre-trained models that can be fine-tuned for specific diagnostic tasks. Each of these techniques contributes to the overall capabilities of AI systems to execute sophisticated image analysis, therefore speeding the diagnostic workflow and enabling physicians to focus on patient care.

In summary, the application of AI in fracture diagnosis and classification represents a significant improvement in orthopedic practice \cite{state}. It offers the potential to not only increase diagnostic accuracy but also to optimize the whole process of fracture therapy, eventually leading to better patient care and results. The continued development and refining of AI technologies will undoubtedly continue to increase their efficacy, establishing their status as a vital component of modern medical practice.

Recent works related to bone fracture classification include the use of CNN, Transformers, etc. In this work, we have focused on the classification of bone fractures from the recent FracAtlas dataset, using transfer learning techniques. Also, we have deployed BAM attention module to enhance the performance of different deep-learning models. As a result of our experimentation, we have achieved a state-of-the-art accuracy of 90.48\%, and a precision of 90.57\%.

In literature review, we have reported the state-of-the-art in bone fracture classification, chapter 3 consists of the detailed explanation of the model used in the experimentation, chapters 4 and 5 reports the experimental setup configurations and experimental results respectively.

\section{Literature Review}

\subsection{Related Works}\label{AA}
Transfer learning models have shown promising results in terms of bone fracture classification. Many transfer learning models such as VGG, Inception, DenseNet, ResNet, etc. have been used in fracture classification and these models provide enough accuracy in many domains. Many researchers used pre-trained models in their work and got promising results. Classification of knee osteoporosis \cite{knee}, is one of the great works on knee osteoporosis classification that showed great performance. In this work, Yang et. al. worked with the Osteoporosis Knee X-ray Dataset \cite{kneedata} and used VGG16 and Late Fushion model on this dataset and got 82\% and 77\% accuracy respectively.  

In many cases of classification of bone fracture, researchers use ensemble approaches to make models more efficient. Sometimes ensemble models provide more efficient results than pre-trained models. KONet \cite{konet}, is a weighted ensemble approach that seamlessly integrates the strengths of EfficientNetB0 and DenseNet121 for knee osteoporosis diagnosis. In this work, this model achieved 97\% accuracy in working with osteoporosis knee X-ray dataset \cite{kneedata} and this work is also recognized as the first ensemble-based approach for knee osteoporosis classification. In many researches, more than one ensemble approach has been used as ensemble models combine the strengths of many pre-trained models. One of these types of research is the classification of shoulder X-ray images \cite{shoulder}. In this paper, Peker et. al. worked with the shoulder section of MURA dataset \cite{mura} and used two ensemble approaches and achieved 84.55\% and 84.72\% accuracy respectively.

Some researchers give more preference to architectural layers to enhance the efficiency of models. They change some methods in architecture and increase proficiency.  Researcher Wang et. al. proposed such a method in their paper Attention mechanism-based deep learning method \cite{gan}. They proposed GAN for obtaining the approximation of manual windowing enhancement. Another researcher Kang et. al. proposed DDA network in their research paper \cite{dda}. They used a DDA layer which aims to make the model learn multiclass features in their proposed DDA network.

In some research, researchers used more than one method to get more accurate results. One of these types of research is the fuzzy rank-based ensemble model \cite{fuzzy}. In this paper, researcher Batra et. al. obtained a dataset from Mendeley \cite{fuzzydata} and used three convolutional neural networks to create an ensemble-based classifier model to work with it. Then they employed a fuzzy rank-based unification of classifiers by taking into account two distinct parameters on the decision scores produced by the aforementioned base classifiers. They achieved 93.5\% accuracy using their proposed ensemble model.
\section{Methodology}

\begin{figure*}[h!]
\begin{center}
\includegraphics[width=\textwidth]{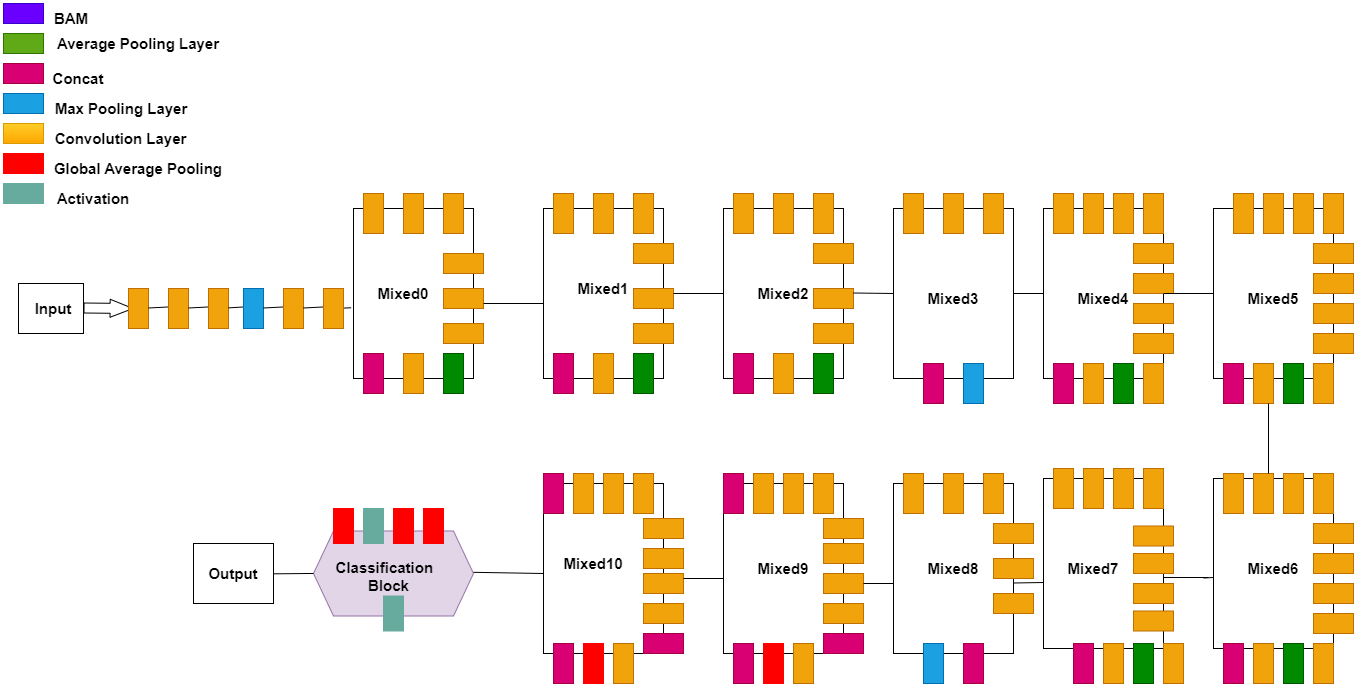}
\end{center}
\caption{InceptionV3 Architecture }
\end{figure*}

\subsection{Dataset description}
There are many datasets on which many researchers have made great achievements. For our research, we have used a new dataset that is FracAtlas \cite{fracatlas} proposed by Abedeen et. al., which was provided by Springer Science and Business Media LLC, and is publicly available via Figshare. It is an X-ray image dataset consisting of 4,083 images manually annotated for classification, localization, and segmentation of bone fractures with the help of 2 expert radiologists and later validated by a medical officer. The data were collected from 3 major hospitals in Bangladesh. There are 717 images with 922 instances of fractures. The rest 3366 images do not contain any instances of fractures and thus are used as the non-fractured class samples. Each of the fracture instances has its own mask and bounding box, whereas the scans also have global labels for classification tasks. The dataset is divided into 2 classes: one is Fractured and the other is Non-fractured.  The dataset consists of images of different aged people’s bones. The age of subjects in the dataset ranges from 8 months to 78 years old. Also, the gender distribution for abnormal studies is 85.4\% and 14.6\% between males and females respectively. The gender ratio for the whole dataset (normal + abnormal cases) is 62\% male and 38\% female approximately.

     

In this dataset, the resolution of all the images are same. The images are all in the same shape. For this dataset, the resolution is 2304×2880. From the fractured and the non-fractured images, two images of leg, hand, and shoulder of each class are demonstrated in Figure \ref{fig:dataset} to show fractures in these parts of the body.

\subsection{Model Architecture}
InceptionV3 \cite{inceptionv3} is used in our research as a backbone model. It is a pre-trained convolutional neural network that is 48 layers deep, which is a version of the network already trained on more than a million images from the ImageNet database. This pre-trained network can classify images into 1000 object categories, such as keyboard, mouse, pencil, and many animals. As a result, the network has learned rich feature representations for a wide range of images. The network has an image input size of 299-by-299. The model extracts general features from input images in the first part and classifies them based on those features in the second part.

\begin{figure}[H]
\begin{center}
\includegraphics[scale=0.35]{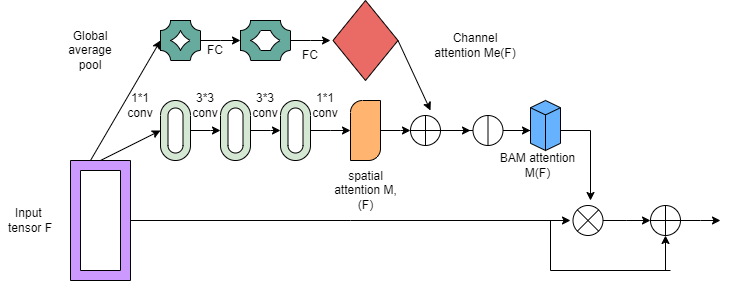}
\end{center}
\caption{BAM Architecture }
\end{figure}

\subsection{BAM Attention Module}
We have chosen BAM as our attention module and added it in our experiment. Bottleneck Attention Module (BAM) \cite{bam} is introduced by Korea Advanced Institute of Science and Technology (KAIST). BAM is a new module that is designed to be integrated with any feed-forward CNNs. This module infers an attention map along two separate pathways, channel and spatial. It is placed at each bottleneck of models where the downsampling of feature maps occurs. BAM constructs a hierarchical attention at bottlenecks with a number of parameters and it is trainable in an end-to-end manner jointly with any feed-forward models. On the CIFAR-100 and ImageNet classification tasks, authors observe performance improvements over baseline networks by placing BAM. The improvement can be observed when multiple BAM modules located at different bottlenecks build a hierarchical attention.

\section{Experimental Setup}

\subsection{Dataset Preprocessing}
We had done three types of work to preprocess the dataset. Firstly, we split the dataset into three subsets: training set, validation set, and test set. we create 3 directories for these 3 sets and split our dataset into those 3 directories. The dataset provider already splits the fracture images into 3 sets. So we just split the non-fractured images into 3 sets. Following splits of fracture images, we split 80\% of non-fractured images into the training set, 8.5\% of non-fracture images into test set, and 11.5\% of non-fracture images into the validation set.

\begin{table}[h!]
    \centering
    \normalsize
    \caption{Split Of Fractured And Non-fractured Classes Into 3 Sets }
    \label{tab:my_label}
   \begin{tabular}{|c|c|c|}
    \hline
    \textbf{Set} &  \textbf{Ratio (\%)} & \textbf{Instances}\\
    \hline
     Training & 80\% & 3266\%\\
    \hline
     Validation & 11.5 & 470 \\
    \hline     
     Testing & 8.5 & 347 \\
     \hline
\end{tabular}
\end{table}

Secondly, we removed the corrupted images from these sets to make the dataset more robust as feeding corrupted images to the model can cause training to halt unexpectedly, leading to wasted time and resources. Lastly, we resized the shape of images from $2304\times2880$ to $224\times224$ as the $224\times224$ picture size is most frequently used as an input size for Convolutional Neural Network (CNN) models. 

\subsection{Experimental Platform}
We have used Google Colab \cite{colab} as experimental platform for our research to get access to GPUs free of charge, and easy sharing. We have used T4 GPU for our research, and got resources of 12.7GB System RAM and 15.0 GB GPU RAM. For every epoch, our runtime required 239ms/step. We have also used different Python libraries to make our work easier. We have used Keras-Application \cite{keras} to import pre-trained models that can be applied to several different applications including image classification and natural language processing, Tenserflow \cite{tenserflow} to build and train models by using the high-level Keras API, and OpenCV \cite{opencv} to work with image processing functions 

\subsection{Hyper-parameter Tuning}
The validation loss has been decreased due to he extensive hyper-parameter tuning that we have performed. Within the scope of our InceptionV3 model architecture, we have added three attention modules. Within the area between the input layer and the output layer, we have added ten custom heads. The following fundamental experimental conditions were set in order to achieve a compromise between the necessity to provide models with sufficient iterations for convergence and the need to maximize the efficiency of computing:

\begin{itemize}
  \item \textbf{Learning Rate:} 1e-3
  \item \textbf{Batch Size:} 32
  \item \textbf{Number of Epochs:} 100
  \item \textbf{Label Mode:} Categorical
  \item \textbf{Optimizer:} Adam
  \item \textbf{Validation Loss Mode:} Min
  \item \textbf{Learning Rate Optimizer:} ReduceLROnPlateau
\end{itemize}

\subsection{Evaluation Metrics}
In this experiment, we have used some popular evaluation metrics. We have used 4 metrics as our evaluation metric. They are Precision, Recall, F1-Score and Accuracy. We have calculated true positives (TP), total samples (TS), false positives (FP), true negatives (TN),  and false negatives (FN) to measure our model's result. The equations for these four metrics are as follows:

\begin{equation}
    \textnormal{Accuracy} =\frac{TN+TP}{TP+FP+TN+FN}
    \label{acc_equ}
\end{equation}

\begin{equation}
    \textnormal{Precision} =\frac{TP}{TP+FP}
\end{equation}

\begin{equation}
    \textnormal{Recall} =\frac{TP}{TP+FN}
\end{equation}

\begin{equation}
    \textnormal{F1-score} =\frac{2\times \textnormal{Precision} \times \textnormal{Recall}}{\textnormal{Precision}+\textnormal{Recall}}
\end{equation}

\section{Experimental Results and Analysis}

\subsection{Results}

\begin{figure*}[h!]
     \centering
     
     \begin{subfigure}[b]{0.49\textwidth}
         \centering
         \includegraphics[width=\textwidth]{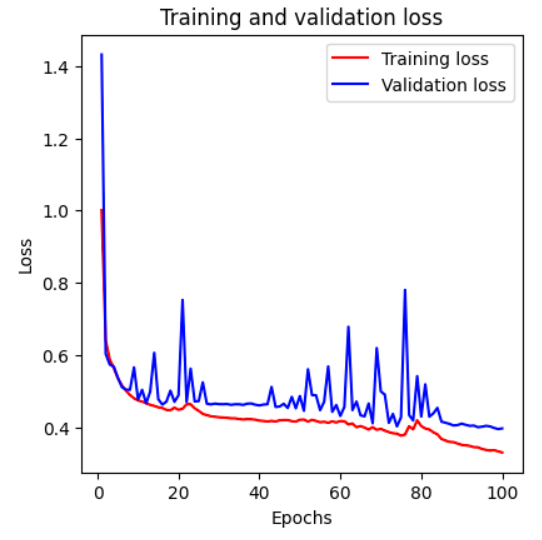}
         \caption{Loss}
         \label{fig:loss}
     \end{subfigure}
     \hfill
     \begin{subfigure}[b]{0.49\textwidth}
         \centering
         \includegraphics[width=\textwidth]{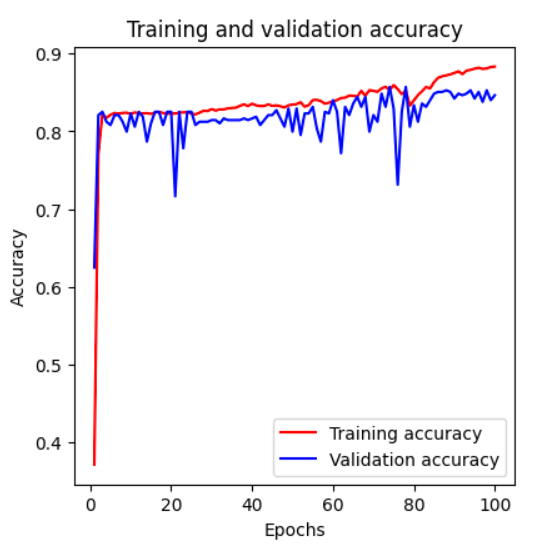}
         \caption{Accuracy}
         \label{fig:acc}
     \end{subfigure}
        \caption{Comparison between Training \& Validation accuracy and loss of Attenuated InceptionV3 Model}
        \label{fig:comparison}
\end{figure*}

The comparison of results was done after doing several experiments using imagenet and without using imagenet. The results from the models are shown in the following tables to see which model has the better result in which section. Here:

\begin{itemize}
    \item Test Accuracy: TA
    \item Test F1 score: TF1
    \item Test Recall: TR
    \item Test precision: TP
\end{itemize}

Table \ref{tab:imagenet_results} contains the results of models using imagenet:

\begin{table}[H]
    \caption{Comparison Of Results Of Models Using Imagenet}
    \label{tab:imagenet_results}
    \centering
    
    \normalsize
    \begin{tabular}{|c|c|c|c|c|}
    \hline
\textbf{Model} & \textbf{TA} & \textbf{TF1} & \textbf{TR} & \textbf{TP} \\
\hline
\textbf{DenseNet121} & \textbf{0.9135} & \textbf{0.9145} & \textbf{0.9145} & \textbf{0.9145} \\
\hline
EfficientNetV2B2 & 0.9048 & 0.9083 & 0.9083 & 0.9083 \\
\hline
InceptionV3  &  0.8991 & 0.9000 & 0.9000 & 0.9000 \\
\hline
EfficientNetB1 & 0.8991 & 0.8985 & 0.8985 & 0.8985 \\
\hline
Inception-v4 & 0.8904 & 0.8901 & 0.8901 & 0.8901 \\
\hline
\end{tabular}
\end{table}

The Table \ref{tab:wo_imagenet} contains results of models without using pre-trained imagenet weights:

\begin{table}[H]
    \caption{Comparison Of Results Of Models Without Using Imagenet}
    \label{tab:wo_imagenet}
    \centering
    \normalsize
    \begin{tabular}{|c|c|c|c|c|}
    \hline
\textbf{Model} & \textbf{TA} & \textbf{TF1} & \textbf{TR} & \textbf{TP} \\
\hline
\textbf{InceptionV3}  &  \textbf{0.9048} & \textbf{0.9057} & \textbf{0.9057} & \textbf{0.9057} \\
\hline
DenseNet121 & 0.8905  & 0.8902  & 0.8902  & 0.8902 \\
\hline
Inception-v4 & 0.8731 & 0.8735 & 0.8735 & 0.8735 \\
\hline
EfficientNetV2B2 & 0.8645 & 0.8651 & 0.8651 & 0.8651 \\
\hline
EfficientNetB1 & 0.8472 & 0.8485 & 0.8485 & 0.8485 \\
\hline
\end{tabular}
\end{table}

\subsection{Qualtitative analysis}
The qualitative analysis of the proposed methodology can be confirmed using several training vs. validation evaluation graphs. From Figure \ref{fig:comparison}, we can see that the graph depicting the propagation of training loss vs. validation loss of the proposed model indicates no case of overfitting or underfitting and the graph showing the propagation of training accuracy vs. validation accuracy of the proposed model indicates that the model achieved a good result.

\subsection{Discussion}
From different experimentations, we got various results and they are compared based on their accuracy, precision, recall, and F1\_score. From Table \ref{tab:imagenet_results} and Table \ref{tab:wo_imagenet}, we can see that DenseneNet121 provides better results using imagenet but InceptionV3 provides better results when we do not use pretrained imagenet weights. Hence, we used InceptionV3 as our backbone model and added three BAM attention modules in its architecture, and also made some hyper-parameter tuning to improve its performance. Before making these experiments, we had seen that our model provided a good accuracy but its graphs showed that the model had overfitting issues. But, after making these experiments, we have seen that though our model's accuracy has declined, it solved our overfitting issue.

\section{Conclusion}
Fracture detection is one of the most challenging problems of bio-medical problems as it directly impacts human life. Fractures in bones can make one’s life challenging and problematic and also cause many hazards in life. So detecting these fractures and identifying their area is always important. In this study, we use the FracAtlas dataset containing a significant number of images. This dataset contains fracture images of different areas and also it shows the normal form of that area. It is a new dataset and this dataset is compatible with classification, segmentation, and localization tasks. This research shows the significant efficiency of InceptionV3 and DenseNet121 deep learning architectures in the domain of X-ray image classification. Also, the application of attention-based mechanism, namely the BAM module, within the transfer learning model enhanced the performance of the architecture quite significantly. This research is about classification tasks and it is a novel work as no classification task has been proposed in this domain using this dataset. We have set a benchmark of achieving 90.48\% accuracy in this domain. This research work also indicates the potential to be deployed with software to automatically detect fractures in the body using X-ray images and is also a testament to much-needed reliability in the domain of medical image analysis.

\bibliographystyle{IEEEtran}

\end{document}